\documentclass[twocolumn,prb]{revtex4}

\usepackage{graphicx}

\begin{document}


\title{Temperature control of local magnetic anisotropy in multiferroic CoFe/BaTiO$_3$} 



\author{Tuomas H. E. Lahtinen and Sebastiaan van Dijken}
\email{sebastiaan.van.dijken@aalto.fi}
\affiliation{NanoSpin, Aalto University School of Science, P.O. Box 15100, FI-00076 Aalto, Finland}


\date{\today}

\begin{abstract}
This paper reports on the temperature evolution of local elastic interactions between ferromagnetic CoFe films and ferroelectric BaTiO$_3$ substrates. Polarization microscopy measurements indicate that growth-induced stripe domains in the CoFe films are preserved and strengthened during cooling and heating through the structural phase transitions of BaTiO$_3$. Moreover, rotation of the magnetic easy axes at the tertragonal-to-orthorhombic transition (T = 278 K) and at T $\approx$ 380 K simultaneously switches the local magnetization of both uniaxial domains by 90\(^\circ\). Irreversible changes in the ferromagnetic domain pattern are induced when the room-temperature ferroelectric domain structure is altered after temperature cycling.  
\end{abstract}


\maketitle 

Multiferroic heterostructures that consist of a thin magnetic film grown onto a ferroelectric substrate have recently been used for studies on magnetoelectric coupling and electric-field controlled magnetism. The magnetic and ferroelectric properties of these hybrid material systems are often indirectly coupled via piezoelectric or ferroelastic strain transfer. For multiferroics driven by piezoelectric strain, the change of magnetization tends to be linear and non-hysteretic in small applied electric field.\cite{thiele_prb_2007,ma_advmater_2011,vaz_electric_2012} Moreover, the induced strain state is uniform across the substrate. Ferroelastic strain transfer from a ferroelectric domain pattern, on the other hand, is laterally modulated. Full imprinting of ferroelectric domains into thin magnetic films has recently been demonstrated.\cite{lahtinen_pattern_2011,lahtinen_electrical_2011,lahtinen:262405,lahtinen_electric-field_2012} If the ferroelectric polarization direction is altered by an electric field, local non-volatile magnetic switching\cite{lahtinen_pattern_2011,lahtinen_electrical_2011} and magnetic domain wall motion can be induced.\cite{lahtinen_electric-field_2012}

One popular approach that has been used to study elastic interactions between a ferroelectric substrate and a magnetic film utilizes the structural phase transitions of BaTiO$_3$ (BTO). The lattice structure of single-crystal BTO substrates changes as a function of temperature from cubic to tetragonal at 393 K, then from tetragonal to orthorhombic at 278 K, and finally from orthorhombic to rhombohedral at 183 K.\cite{Kay_1949} The concurrent change of the BTO domain pattern at these phase transitions alters the strain state of the magnetic film and via inverse magnetostriction this can lead to local magnetic switching. The macroscopic response of several materials including La$_{1-x}$Sr$_{x}$MnO$_{3}$,\cite{lee_apl_2000,eerenstein_giant_2007} Fe$_{3}$O$_{4}$,\cite{tian_apl_2008,vaz_apl_2009,sterbinsky_apl_2010} and Fe\cite{shirahata_switching_2011,brivio_apl_2011,venkataiah_strain-induced_2012} on BTO has been monitored as a function of temperature using SQUID magnetometry, vibrating sample magnetometry (VSM), and magneto-optical Kerr effect measurements. Abrupt changes in sample magnetization at the phase transitions of BTO are a common feature in such experiments, illustrating significant elastic coupling between the ferroelectric substrate and the magnetic film. Macroscopic measurements, however, only reveal an average magnetic response. Due to the strong local character of ferroelastic strain transfer, the change in magnetization varies from one domain to the other. Moreover, a variety of ferroelectric domain transformations can occur at the BTO phase transitions, which complicates the interpretation of macroscopic data. For a full microscopic analysis of ferroelastic strain transfer from BTO substrates to magnetic films, ferroelectric and ferromagnetic domain imaging and local magnetic measurements as a function of temperature are required.

\begin{figure} [b]
\includegraphics{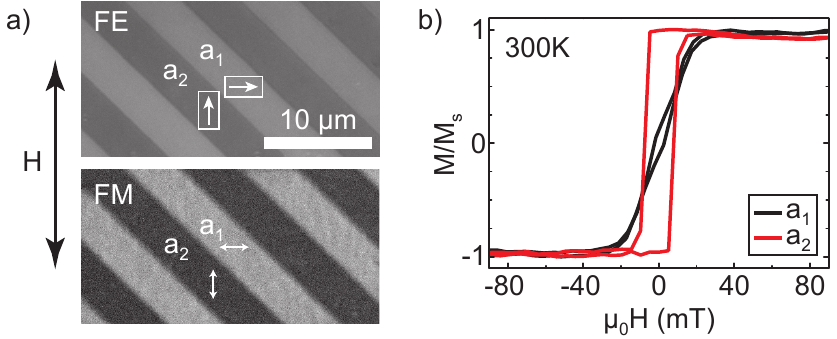}
\caption{(a) Polarization microscopy images of the ferroelectric (FE) and ferromagnetic (FM) domain structures at 300 K. The arrows indicate the polarization direction in the BTO substrate and the orientation of the uniaxial magnetic easy axes in the CoFe film. (b) Local magnetic hysteresis curves of $a_1$ and $a_2$ stripe domains. The magnetic field axis in these measurements is indicated by \(H\)}
\end{figure}

Here, we present a study on domain pattern transfer and local magnetic anisotropy in multiferroic CoFe/BTO. Using a polarization microscope with an integrated cryostat, we imaged and analyzed the ferroelectric and ferromagnetic domains in a temperature range spanning four structural phases of BTO. The experiments reveal (1) that the ferroelectric domain pattern of BTO substrates is fully imprinted into CoFe during film growth at 300 K, (2) that this pattern is preserved and the magnetoelastic anisotropy is considerably enhanced upon cooling and heating through the BTO phase transitions, and (3) that cycling back to 300 K irreversibly changes the ferromagnetic domain pattern of CoFe films on tetragonal BTO.

\begin{figure}
\includegraphics{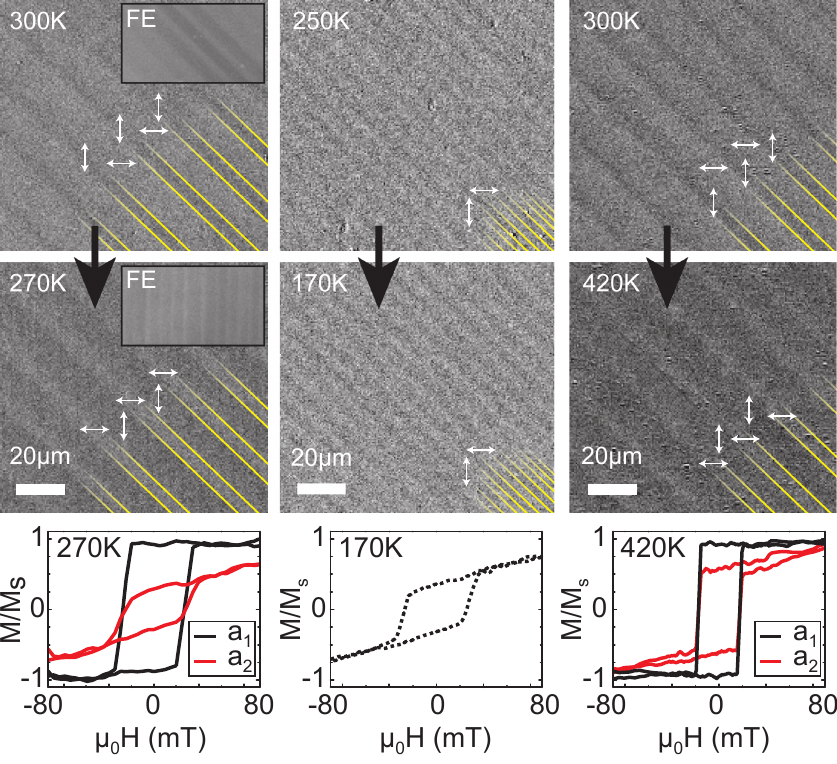}
\caption{Images of the CoFe domain structure for different temperatures and zero applied magnetic field. The magnetic stripe pattern of the CoFe film is preserved on all structural phases of the BTO substrate. The magnetic contrast of the stripe domains reverses when the BTO substrate is cooled from 300 K to 270 K or heated from 300 K to 420 K, indicating 90\(^\circ\) magnetic switching within the domains. A similar effect does not occur when the sample is cooled through the orthorhombic-to-rhombohedral phase transition (250 K $\rightarrow$ 170 K). Ferroelectric domain images of the BTO substrate at 300 K and 270 K are shown in the insets. We note that the three columns of images were obtained in separate experiments. The differences in domain stripe width are not due to a temperature effect, but caused by domain variations across the BTO substrates. Moreover, the lower resolution and contrast of these images compared to Fig. 1 are caused by the cryostat. The graphs show magnetic hysteresis curves on single $a_1$ and $a_2$ stripe domains measured at 270 K and 420 K. The curve for 170 K is a macroscopic measurement, i.e. an average over many $a_1$ and $a_2$ domains.}
\end{figure}

CoFe films with a composition of 60$\%$ Co and 40$\%$ Fe were grown onto tetragonal BTO substrates with a regular ferroelastic $a_1-a_2$ domain pattern using electron-beam evaporation at 300 K. The thickness of the CoFe films was 15 nm and a 3 nm Au capping layer was used to prevent oxidation. The metallic layers were semi-transparent for white light, which enabled simultaneous optical imaging of the ferroelectric domains in the BTO substrate and of the ferromagnetic domains in the CoFe film. Contrast from the domain patterns was obtained via birefringence (ferroelectric domains) and the magneto-optical Kerr effect (ferromagnetic domains) using a modified polarization microscope. For variable temperature measurements, the samples were inserted into a microscope cryostat. Local magnetic hysteresis curves were measured on single domains by the acquisition of magneto-optical Kerr effect data from pre-selected sample locations. The magnetic field was aligned at an angle of 45\(^\circ\) with respect to the domain stripe pattern as indicated in Fig. 1.    

The $a_1-a_2$ domain pattern of the tetragonal BTO substrates is characterized by 90\(^\circ\) rotations of the ferroelectric polarization and tetragonal lattice elongation in the substrate plane. At a deposition temperature of 300 K, this produces an uniaxial strain modulation of 1.1\% (a = 4.035 \AA, b = c = 3.991 \AA).\cite{kwei_structures_1993} Figure 1 shows domain images and local magnetic hysteresis curves after CoFe film growth. The data demonstrate that the BTO domains are imprinted into the CoFe film via local strain transfer and inverse magnetostriction. The strain-induced uniaxial magnetic easy axes in the $a_1$ and $a_2$ domains are orthogonal as illustrated by the hard- and easy-axis hysteresis curves. From the slope of the hard-axis curve, the magnetoelastic anisotropy is determined as $K_{me}=1.7\times10^4$ J/m$^3$. This anisotropy is considerably smaller than $K_{me,max}=2.8\times 10^5$ J/m$^3$,\cite{PhysRevB.85.094423} which is the estimated maximum anisotropy based on full 1.1\% strain transfer from the BTO substrate. Our experiments thus indicate that most of the BTO lattice strain is relaxed during CoFe film growth at 300 K.

\begin{figure}
\includegraphics{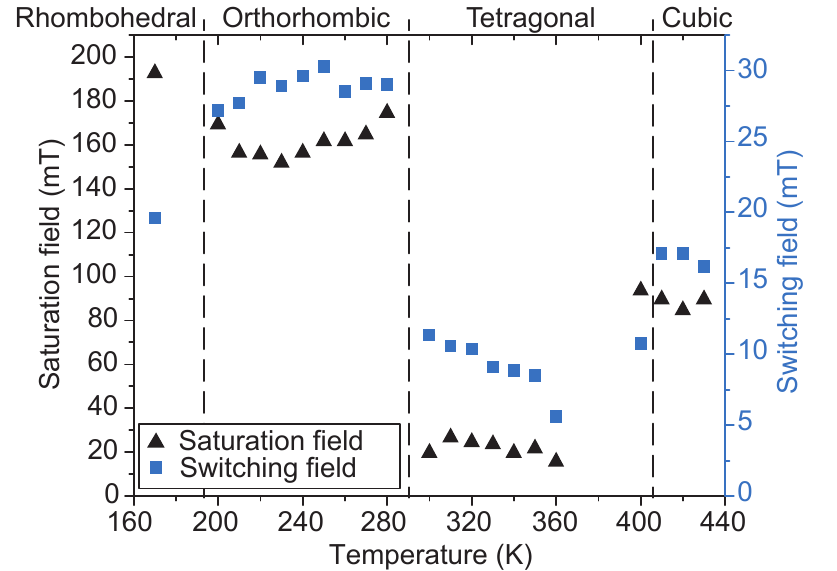}
\caption{Saturation field of local hard-axis magnetization curves (black triangles) and switching field of local easy-axis hysteresis curves (blue squares) as a function of temperature.}
\end{figure}

\begin{figure}
\includegraphics{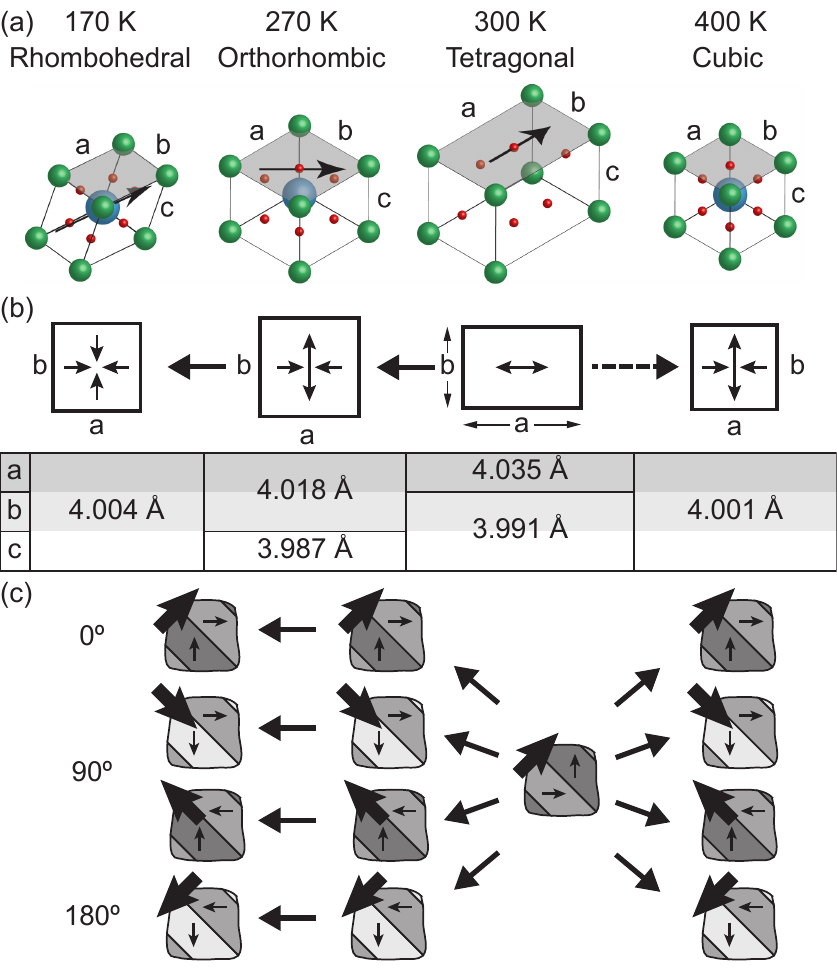}
\caption{(a) Schematic illustration of the BTO unit cell for different structural phases. The gray side is oriented in the BTO substrate plane in our experiments. The arrows indicate the direction of ferroelectric polarization. (b) BTO lattice structure in the substrate plane and BTO lattice parameters for T = 170 K, T = 270 K, T = 300 K, and T = 400 K.\cite{kwei_structures_1993,megaw_temperature_1947} The arrows indicate the direction of uniaxial strain transfer to the CoFe films at the BTO phase transitions. (c) Illustration of the magnetic CoFe domain pattern for different structural phases of BTO. The small arrows represent the local magnetization direction and the big arrows indicate the macroscopic sample magnetization which, depending on the direction of local magnetization rotation, switches by 0\(^\circ\), 90\(^\circ\), or 180\(^\circ\).}
\end{figure}

After micromagnetic analysis at room temperature, the samples were cooled and heated through the structural phase transitions of the BTO substrates. Upon cooling, the BTO lattice changes from tetragonal to orthorhombic at 278 K and subsequently to rhombohedral at 183 K.\cite{Kay_1949} The BTO phase transitions are accompanied by abrupt modifications of the ferroelectric domain pattern and, thus, the exertion of new local strains on the CoFe film. Similar effects do occur when the samples are heated through the tetragonal-to-cubic phase transition at 393 K. Despite the variation of transferred ferroelastic strain, Kerr microscopy measurements indicate that the growth-induced magnetic stripe pattern is preserved upon cooling and heating through all BTO phase transitions (Fig. 2). Qualitatively, this behavior can only be explained by the absence of strong uniaxial in-plane lattice strains in the rhombohedral, orthorhombic, and cubic phases of the BTO substrates, i.e. by cubic in-plane structural symmetry. To study the effects of local strain transfer, magnetic hysteresis curves were recorded on individual $a_1$ and $a_2$ domains as a function of temperature. A representative selection is shown in Fig. 2 and the temperature dependence of the CoFe saturation and switching fields are summarized in Fig. 3. Hereafter, we will discuss the temperature evolution of the local CoFe film properties for each of the BTO phases.

During the BTO phase transition at 278 K, the ferroelectric polarization switches from an elongated $<$100$>$ direction in the tetragonal phase to a $<$110$>$ direction along the diagonal of the unit cell plane that is formed by the two largest lattice parameters in the orthorhombic phase (a = b = 4.018 \AA, c = 3.987 \AA).\cite{kwei_structures_1993,forsbergh_domain_1949} The polarization microscopy images in the insets of Fig. 2 confirm a 45\(^\circ\) rotation of the ferroelectric domains. The growth-induced magnetic stripe pattern of the CoFe film, however, is entirely preserved during the phase transition. These observations indicate that the in-plane lattice strain of neighboring ferroelectric domains in the orthorhombic phase is identical in our samples. In other words, the $a$ and $b$ lattice parameters are oriented parallel to the BTO substrate plane as schematically illustrated in Fig. 4. While the magnetic domain pattern is preserved, the local switching and saturation fields of the CoFe film increase sharply when the BTO structure transforms from tetragonal to orthorhombic (Fig. 3). These effects can be explained by considering the abrupt changes in the BTO lattice. Upon cooling through the phase transition at 278 K, the BTO lattice is compressed along $a$ and elongated along $b$. This transformation completely removes the in-plane 1.1$\%$ uniaxial lattice strain of the tetragonal BTO substrate. If the efficiency of ferroelastic strain transfer to the CoFe film would be identical during growth and sample cooling, this would terminate the uniaxial magnetoelastic anisotropy. However, because the CoFe film is clamped to the BTO substrate, strain transfer is more efficient during temperature cycling. As a result, the uniaxial strain direction and the magnetoelastic easy axes of the CoFe film rotate by 90\(^\circ\) at the tetragonal-to-orthorhombic phase transition, i.e. the magnetic $a_1$ domains start to behave like $a_2$ domains and vice versa. For the  $a_1$ domains, this is clearly illustrated by a change in the shape of their hysteresis curve from hard-axis at 300 K (Fig. 1) to easy-axis at 270 K (Fig. 2). An opposite effect is measured on the $a_2$ domains. The local magnetization thus abruptly rotates by 90\(^\circ\) when the sample is cooled in zero applied magnetic field. This effect is further confirmed by a reversal of magnetic domain contrast in the images of Fig. 2 (300 K $\rightarrow$ 270 K). From the increase of the saturation field from about 20 mT in the tetragonal phase to about 170 mT in the orthorhombic phase (Fig. 3), an enhancement of the magnetoelastic anisotropy to  $K_{me}\approx1.4\times10^5$ J/m$^3$ is estimated. This anisotropy is of the same order as $K_{me,max}$, which confirms that strain transfer from the BTO substrate to the CoFe film is indeed more efficient during sample cooling than during growth. 

We note that preservation of the magnetic domain pattern in combination with rotation of the uniaxial magnetic easy axes within the domains can alter the macroscopic sample magnetization in different ways. If the magnetization of neighboring stripe domains rotate in the same direction, the macroscopic sample magnetization also rotates by 90\(^\circ\). Macroscopic magnetization reversal by 0\(^\circ\) or 180\(^\circ\) is, however, obtained when the magnetization of the $a_1$ and $a_2$ domains switch in opposite directions. These configurations are schematically illustrated in Fig. 4(c). Thus, the macroscopic magnetic properties of multiferroic heterostructures, as often measured using SQUID or VSM techniques, do not only depend on the type of ferroelectric and magnetic domains that form at BTO phase transitions, but also on local magnetization reversal trajectories towards new local energy minima. The large variety in macroscopic magnetization effects at the BTO phase transitions ($\Delta$$M/M$), which can be found in literature,\cite{lee_apl_2000,eerenstein_giant_2007,tian_apl_2008,vaz_apl_2009,sterbinsky_apl_2010,shirahata_switching_2011,brivio_apl_2011,venkataiah_strain-induced_2012} are likely caused by a difference in the chirality of local magnetic switching. 

The BTO substrates undergo an orthorhombic-to-rhombohedral phase transition at 183 K. The ferroelectric polarization in the rhombohedral phase points along one of the $<$111$>$ directions and the BTO lattice parameters are identical (a = b = c = 4.004 \AA).\cite{kwei_structures_1993} Due to four-fold structural symmetry, the magnetic domain pattern of the CoFe film is again preserved (Fig. 2). Compared to the orthorhombic phase, the compressive strain along the $a$ axis is enhanced and the tensile strain along the $b$ axis is reduced. Because both strain effects almost cancel each other in terms of magnetostriction, the magnetoelastic anisotropy changes only slightly. This is illustrated by the relative small adjustments of the saturation and switching fields below 183 K (Fig. 3). 

Upon heating from 300 K, the BTO substrates undergo a phase transition at the ferroelectric Curie temperature of 393 K. Above this temperature, BTO is cubic with a = b = c = 4.001 \AA.\cite{megaw_temperature_1947} Unlike the abrupt BTO phase transitions at low temperatures, the tetragonal-to-cubic transition is more gradual. As a result, the BTO lattice elongation reduces continuously with increasing temperature, thereby opposing the growth-induced strains of the CoFe films. Because strain transfer is more efficient during sample heating than during CoFe film growth, the original CoFe strain state is already fully compensated below the BTO Curie temperature. The experimental data of Fig. 3 confirm this. Initially the saturation and switching fields decrease due to thermal activation. However, both parameters increase sharply at about T $\approx$ 380 K indicating an enhancement of magnetoelastic anisotropy. In addition, contrast reversal in the polarization microscopy images and a change in the shape of local magnetic hysteresis curves (Fig. 2) reveal that the uniaxial magnetic easy axes rotate by 90\(^\circ\). Since the BTO lattice is cubic above 393 K, the magnetic domain pattern is again preserved and the macroscopic magnetization either reverses by 0\(^\circ\), 90\(^\circ\), or 180\(^\circ\) depending on the chirality of local magnetic switching.   

\begin{figure}
\includegraphics{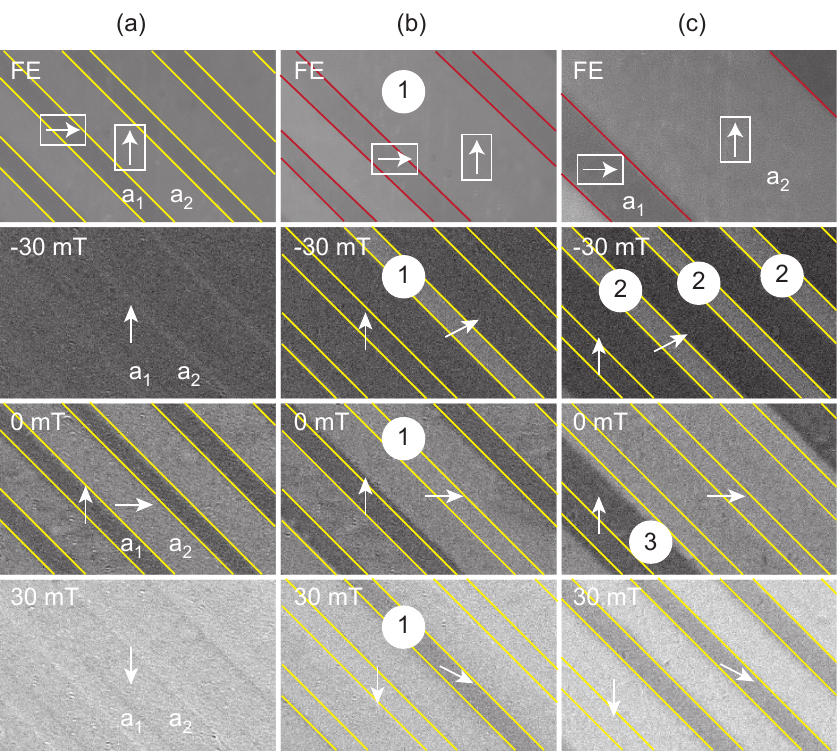}
\caption{Polarization microscopy images of the ferroelectric (FE) and ferromagnetic domain structure at 300 K (tetragonal BTO). The images represent (a) the as-deposited state, (b) after one temperature cycle to orthorhombic BTO, and (c) after a second temperature cycle. The yellow lines in the images indicate the domain pattern of the BTO substrate during CoFe film growth. The red lines in (b) and (c) illustrate the modified BTO domain structure after temperature cycling. The arrows indicate the polarization direction in the BTO substrate and the direction of magnetization in the CoFe film. The numbers label the areas in which the in-plane polarization of the BTO substrate rotates by 90\(^\circ\).}
\end{figure}

Finally, we analyze the evolution of the magnetic domain pattern upon repeated temperature cycling between the tetragonal and orthorhombic phases of BTO. Figure 5(a) shows the as-deposited ferroelectric (FE) and magnetic domain structure for different magnetic field. In this state, the CoFe magnetization is fully saturated at -30 mT because of the modest magnetoelastic anisotropy of $K_{me}=1.7\times10^4$ J/m$^3$. Coherent magnetization rotation towards the uniaxial magnetic easy axes of the $a_1$ and $a_2$ domains magnetically reproduces the ferroelectric stripe pattern at small magnetic field. A reversal and increase of the applied field to +30 mT saturates the magnetization along the new field direction. Figure 5(b) shows the same sample area after cooling to the othorhombic BTO phase and heating back to tetragonal BTO at 300 K. During temperature cycling across the BTO phase transition, one of the ferroelectric $a_1$ domain stripes was assimilated by neighboring $a_2$ domains (indicate by (1)). On top of the reversed ferroelectric domain, the CoFe magnetization no longer saturates in an applied field of 30 mT because of a significant increase in magnetoelastic anisotropy. The saturation field of the new magnetic stripe domain amounts to about 600 mT, which corresponds to $K_{me}\approx5.1\times10^5$ J/m$^3$. This anisotropy value almost equals 2$K_{me,max}$, which is expected for full ferroelastic strain transfer from an $a_1$ to $a_2$ domain transition in the BTO substrate (1.1\% lattice compression plus 1.1\% lattice elongation in two orthogonal directions). During a second temperature cycle, two more $a_1$ domains turned into $a_2$ stripes (indicate by (2)) and one  $a_1$ domain grew in size (indicate by (3) in Fig. 5(c)). The local magnetic response of the CoFe film is the same, i.e. an increase and 90\(^\circ\) rotation of the magnetic anisotropy in areas where ferroelectric $a_1$ domains are replaced by $a_2$ domains and vice versa or, in other words, in areas where the in-plane ferroelectric polarization is rotated by 90\(^\circ\). 

\begin{figure}
\includegraphics{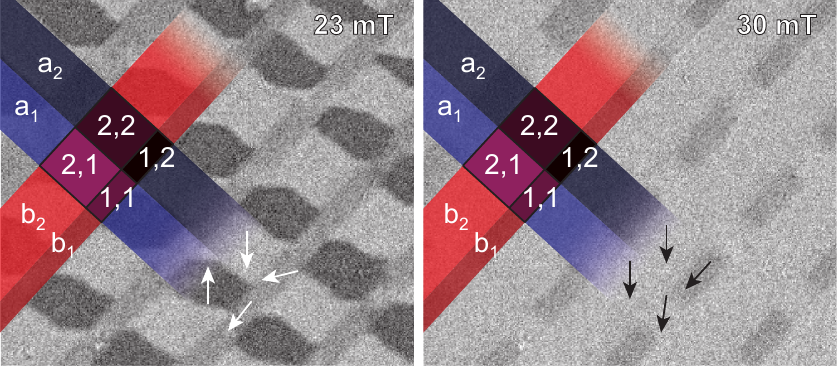}
\caption{Magnetic domain images of the CoFe film at 300 K for two magnetic fields. The images were recorded after one temperature cycle to orthorhombic BTO. The $a_1$ and $a_2$ stripes indicate the growth-induced domain pattern and $b_1$ and $b_2$ refer to stripe domains that are formed during temperature cycling. Four magnetic anisotropy states exist. The $b$,$a$ labels contain information about the orientation of the uniaxial magnetic easy axis ($b=1$ horizontal, $b=2$ vertical) and the strength of magnetic anisotropy ($b=a$ small growth-induced anisotropy, $b$ $\neq$ $a$ large temperature-induced anisotropy)}
\end{figure}

Another example of a temperature-induced magnetic domain structure is shown in Fig. 6. In this case, the ferroelectric $a_1-a_2$ stripe pattern of the BTO substrate rotated by 90\(^\circ\) during cooling and subsequent heating across the tetragonal-to-orthorhombic phase transition. Rotation of the BTO domains is possible because tetragonal stripe patterns along [110] and [-110] are energetically equivalent. The strain-driven response of the CoFe film is characterized by a superposition of the as-deposited magnetic domain structure (labeled by $a_1$ and $a_2$ in Fig. 6) and temperature-induced domain stripes ($b_1$ and $b_2$). Since strain transfer is more efficient during temperature cycling than CoFe film growth, the magnetic anisotropy axes are determined by the uniaxial strain of the $b$ domains. The new magnetic configuration contains four areas with different anisotropy (two anisotropy orientations and two anisotropy strengths), which produce a checkerboard pattern when a magnetic field is applied.       

In conclusion, we have demonstrated that local magnetic properties of CoFe films on BTO substrates change significantly at the structural phase transitions of BTO. Starting from the as-deposited state at 300 K, cooling through the tetragonal-to-orthorhombic phase transition results in abrupt 90\(^\circ\) magnetic switching in magnetic stripe domains, an increase of the magnetoelastic anisotropy, and the conservation of the overall magnetic domain pattern. Similar effects are observed upon heating towards the tetragonal-to-cubic phase transition. A return to tetragonal BTO, irreversibly changes the magnetic domain structure of the CoFe film. The local magnetic response fully correlates with the temperature-induced assimilation, growth, or rotation of ferroelectric domains in the BTO substrate.



%
%

%

\begin{acknowledgments}
This work was supported by the Academy of Finland (Grant Nos. 127731 and 260361) and the European Research Council (ERC-2012-StG 307502-E-CONTROL). T.H.E.L. acknowledges support from the National Doctoral Program in Materials Physics. 
\end{acknowledgments}


\end{document}